\documentclass[aps,prl,a4paper,twocolumn,showpacs,showkeys,preprintnumbers,amsmath,amssymb]{revtex4}
\usepackage{graphicx}
\usepackage{bm}
\usepackage{dcolumn}
\usepackage{color}
\usepackage{pst-plot}

\newcommand{\be}{\begin{equation}}
\newcommand{\ee}{\end{equation}}
\newcommand{\bea}{\begin{eqnarray}}
\newcommand{\eea}{\end{eqnarray}}

\begin{document}

\title{A minimal length versus the Unruh effect}

\author{Piero Nicolini}
\thanks{Electronic address: nicolini@th.physik.uni-frankfurt.de}
\affiliation{Frankfurt Institute for Advanced Studies (FIAS),
Institut f\"ur Theoretische Physik, Johann Wolfgang Goethe-Universit\"at, 
Ruth-Moufang-Strasse 1, 60438 Frankfurt am Main, Germany}
\author{Massimiliano Rinaldi}
\thanks{Electronic address: massimiliano.rinaldi@unige.ch}
\affiliation{D\'epartment de Physique Th\'eorique, Universit\'e de Gen\`eve,
24 quai Ernest Ansermet
  CH--1211 Gen\`eve 4, Switzerland.}

\date{\today}

\begin{abstract}
\noindent In this Letter we study the radiation measured by an accelerated detector, coupled to a scalar field, in the presence of a fundamental minimal length. The latter is implemented by means of a modified momentum space Green's function. After calibrating the detector, we find that the net flux of field quanta is negligible, and that there is no Planckian spectrum. We discuss possible interpretations of this result, and  we comment on experimental implications in heavy ion collisions and atomic systems.
\end{abstract}


\maketitle

\noindent The emergence of a minimal length in a quantum spacetime is an old idea, dating back to the early times of Quantum Gravity \cite{DW}. In most cases,  it turns out to be the crucial signature in every phenomenon that takes place on a background that departs from a purely classical description. In this general framework, the study of the Unruh effect in the presence of a minimal length can lead both to profound insights and simple phenomenological predictions. In fact, acceleration radiation has a prominent role in a variety of physical contexts: beyond the theoretical case of an accelerated detector, the Unruh radiation might affect the transverse polarization of electrons and positrons in particle storage rings (Sokolov–Ternov effect) \cite{Akhmedov:2006nd,Akhmedov:2007xu}, and the onset of the Quark Gluon Plasma (QGP) due to heavy ions collisions \cite{Kharzeev:2005iz}. The Unruh effect might have non-negligible imprints in low energy physics too, such as the dynamics of electrons in Penning traps, of atoms in microwave cavities, and of ultraintense lasers (for a review see Ref.\ \cite{Crispino:2007eb} and references therein). Finally, its companion effect, {\rm i.e.} the Hawking radiation, is extensively investigated in analog models of gravity, such as Bose-Einstein condensates (BEC) \cite{Barcelo,Balbinot,Carusotto:2008ep}.

The presence of a minimal length $\ell$ is testable only if one can perform experiments at energies around the scale $M_*=1/\ell$. However, we recall that  low energy systems are also endowed with relevant microscopic scales whose global effects, though important, cannot be described by the larger scale effective models often in use.
On the other hand, fine tuning experiments in condensed matter systems and very high energy particle collisions are now in progress and could reveal key information about the interplay between the Unruh effect and the existence of a coarse-grained background in the system \cite{rosu}. It is therefore imperative to have an accurate description of the acceleration radiation in the presence of a minimal length.

The energy scale associated with a minimal length is typically seen as the frontier beyond which local Lorentz symmetry is violated, and it is usually set to be of the order of the Planck mass, as in the vector-tensor theories of gravitation \cite{Jacob}. In other cases, such as in analog models in BEC \cite{Barcelo,Balbinot}, this energy scale is much smaller. In both contexts the violation appears as a modification of the dispersion relation. This possibility was widely studied in relation to the transplanckian problem in cosmology (see e.g. \cite{Martin,staro}), and to the robustness of both Hawking emission \cite{Corley,Unruh} and Unruh effect \cite{MaxUnruh}. The lesson learnt from these works is that the minimal length associated with modified dispersion relations has a negligible impact on these phenomena.

The acceleration radiation was also studied in the case when the minimal length is introduced to cure the divergent ultraviolet (UV) behaviour of the field theory. For example, in \cite{PaddyDual} the propagator is modified via path integral duality, and it is finite in the UV regime. In \cite{Pepe1,Pepe2}, the same propagator is found by deforming the action of the Lorentz group. As for modified dispersion relations, the effect  on both the Unruh effect  \cite{Paddy,Pepe1,Pepe2} and on the Hawking radiation \cite{Pepe3} is negligible.

Lorentz-violating models are increasingly disfavored by observations, see e.g. \cite{Granot}. Therefore, it seems more sensible to implement a Lorentz invariant length $\ell$ in the theory. In the following, we do not assume any particular value for $\ell$, which presumably depends on the details of the underlying quantum gravitational theory. A natural choice would be a value of the order of the Planck length ($10^{-35}$ m), which could be larger in the presence of extra-dimensions, as discussed at the end of the Letter.  

With this spirit in mind, in this Letter we assume that the Euclidean momentum space propagator is given by \cite{Smailagic:2003rp,Smailagic:2003yb,Smailagic:2004yy}
\bea
G_\ell(p^2)=\frac{e^{-\ell^2 p^2/2}}{ p^2+m^2}\ ,
\label{ginp}
\eea
where $p^2=p_0^2+|\vec{p}|^2$.
A similar propagator was already successfully employed in the context of both black hole physics \cite{Nicolini:2005zi,Nicolini:2005vd,Ansoldi:2006vg,Rizzo:2006zb,Spallucci:2008ez,Nicolini:2009gw} (for a review see Ref. \cite{Nicolini:2008aj} and further references therein), and of inflationary cosmology \cite{Rinaldi:2009ba}. The main result is that the divergent short distance behavior of the conventional solutions to field equations (including the ones on curved spacetimes) is cured while, as expected, the quantum fluctuations of the manifold do not occur at scales larger than $\ell$, where the classical description of gravity efficiently works. In particular, 	
the divergent behavior of the black hole evaporation in the Planck phase has been regularized. In the new scenario, the terminal stage of the Hawking quantum emission is in fact characterized by a thermodynamically stable (positive heat capacity) phase of cooling  of the black hole, often called the ``SCRAM phase'' \cite{Spallucci:2006zj,Casadio:2008qy}.

We begin our discussion by briefly recalling the main features of the Unruh effect, as presented in Ref. \cite{BirDav}. We consider a detector, moving in a flat background spacetime along a trajectory $x^\alpha (\tau)$, where $\tau$ is the detector proper time. We assume that the detector moves through a region permeated by a quantum scalar field $\phi$, and that the interaction between the two can be described in terms of the Lagrangian $L_{\mathrm {int}}=\gamma \ \mu (\tau)\ \phi[x^\alpha (\tau)]$, where $\gamma$ is a small coupling constant and $\mu$ is the detector monopole momentum operator. Due to the interaction with the field, the detector will undergo a transition from the ground state $E_0$ to  an excited state $E>E_0$. As $\gamma$ is small, we can derive the transition probability $\Gamma=\int dE \ |\Psi|^2$ by squaring the first order amplitude
\be
\Psi=i\langle E\,; \psi|\int_{-\infty}^\infty L_{\mathrm{int}}\ d\tau\ |0_{\mathrm{M}};\, E_0\rangle\ ,
\ee
where $|0_{\mathrm{M}} \rangle$ is the Minkowski vacuum, and $|\psi \rangle$ is the field excited state. At the lowest order, the monopole operator is well approximated by $\mu (\tau)= e^{iH_0\tau}\mu (0)e^{-iH_0\tau}$, hence we can separate the contributions of the detector and the field to the amplitude by writing
\bea
\Psi=i\gamma\langle E\,|\mu (0)|\,E_0\rangle\int_{-\infty}^\infty  d\tau e^{i\Delta E\tau}\langle\psi|\phi(x)|0_{M}\rangle\ ,\label{sepcont}
\eea
where $\Delta E=E-E_0$. From this one sees that, at first order, the state $|\psi\rangle$ can only contain a single field quantum. However, to find the transition probability, we need to take in account transitions to all possible energies, thus
\be
\Gamma=\gamma^2\sum_E |< E |\mu (0) | E_0>|^2\,{\cal F}(\Delta E)\ ,
\ee
where the detector response function ${\cal F}(\Delta E)$ is given by
\be
{\cal F}(\Delta E)=\int_{-\infty}^\infty d\tau\int_{-\infty}^\infty d\tau^\prime e^{-i\Delta\tau\Delta E} G^+ \left(x(\tau), x(\tau')\right).
\label{respfunction}
\ee
Here , $\Delta\tau=\tau-\tau^\prime$, and $G^+$ is the positive frequency Wightman-Green function.   We stress that the response function is fully specified in terms of the properties of the field, and it does not depend on the choice of the detector, whose sensitivity is given only by $S=\gamma^2\sum |\langle E |\mu (0) | E_0\rangle|^2$. The double integration in Eq.\ (\ref{respfunction}) means that the flux of particles interacting with the detector diverges as soon as the detector-field system  reaches an equilibrium configuration. Therefore, one usually considers the transition probability per unit proper time, $\dot \Gamma=S \dot{\cal F}$, where we define the response rate
\be\label{normresp}
\dot {\cal F}=\int_{-\infty}^\infty d\Delta\tau e^{-i\Delta\tau\Delta E} G^+ (\Delta x)\ .
\ee
In this expression, $\Delta x^2=\eta_{\mu\nu}(x^{\mu}-x'^{\mu})(x^{\nu}-x'^{\nu})$ is the Minkowski proper time interval squared.
For an inertial detector moving with constant velocity $v$, one has $\Delta x^2=\Delta\tau^2/(1-v^2)$,  and $G^+ (\Delta x) $ diverges when $\Delta\tau\to 0$.   However, as no other singularities occur on the integration path, one can show that $\dot{\cal F}$ vanishes by means of the $i\epsilon$ prescription. On the contrary, when the trajectory is not inertial,
the Minkowski interval has the form $\Delta x^2=f(\Delta\tau)$, where $f$ is a non-constant and finite function.
Therefore,
the integrand function in (\ref{normresp}) exhibits poles corresponding to each zero of  $f(\Delta\tau)$ and the rate is no longer vanishing. For example, for a uniformly accelerated detector, with acceleration $1/\alpha$,  coupled to a massless scalar field, one finds a non-vanishing rate $\dot {\cal F}\sim \exp(-2\pi\alpha\Delta E)$. Thus, we learn that the detector feels an incoming radiation of quanta, as if it was coupled to a thermal bath at the temperature $T=1/2\pi \alpha k_B$ \cite{UnruhEff}. 

The above calculations can also be performed in Euclidean space, upon the analytic continuation $i\tau=\tau_E$. Then, the response rate formula becomes
\be
\dot {\cal F}=i\int_{i\infty}^{-i\infty} d\Delta\tau_E e^{\Delta \tau_E\Delta E} G^+ _E(\Delta x)\ ,
\label{resprate}
\ee
where $G^+ _E$ is the Euclidean Wightman function. A detector with uniform acceleration $1/\alpha$ on the Euclidean plane typically  follows a circular trajectory of the form $\alpha^2 \sin^2\left(\Delta \tau_E/ 2\alpha\right)$. Below we will find more convenient to work in Euclidean space, thus we will use Eq.\ (\ref{resprate}), instead of (\ref{normresp}) to calculate the radiation flux.

We now proceed with the  implementation of a minimal length in the framework of the  Unruh effect, by adopting the propagator (\ref{ginp}). We see that the minimal length appears in the damping factor, and this is physically interpreted as a blurring, or delocalization, occurring at each point on a manifold when probed by high momenta. However, at lower momenta the presence of $\ell$ is actually negligible and, usually, one can work with the ordinary field theory.
The Euclidean propagator in coordinate space can be found by calculating the Fourier transform of the Schwinger representation
\bea
 \frac{e^{-\ell^2 p^2/2}}{p^2+m^2}=e^{\ell^2 m^2/2}\int_{\ell^2/2}^{\infty}ds\,e^{-s(p^2+m^2)}.
 \eea
In the massless case, we find that the modified Euclidean Wightman-Green function is \cite{MaxNew}
\be\label{Prop}
G_\ell^E(\Delta x)=-\frac{1}{ 4\pi^2 (\Delta \vec{x\,}^2+\Delta t^2_E)}\left[1-e^{-(\Delta t^2_E+\Delta \vec{x\,}^2)/2\ell^2}\right]\ .
\ee
The theory behaves nicely, as $G_\ell^E$ reduces to its conventional form in the limit $\ell\rightarrow 0$.
More importantly,  the above function  shows its regularity at coincident points: in the double limit $(\Delta t_E, \Delta \vec{x\,}^2)\rightarrow (0,0)$ one has  $G_\ell^E\rightarrow -1/8\pi^2\ell^2$.
The same holds for the massive case, as one can show that, in the coincidence limit,
\be\label{Propm}
G_\ell^E\sim -\frac{1}{8\pi^2\ell^2}+ m^2e^{m^2\ell^2/2} E_1(m^2\ell^2/2)\ ,
\ee
where the exponential integral $E_1(m^2\ell^2/2)$ is finite, and vanishes smoothly in the massless case.

We now calculate the the radiation seen by a detector moving with constant acceleration $1/\alpha$. Note that, at first order in perturbation theory, we can still consider the detector as an ideal point-like object, while considering delocalization as affecting  the field only. If the trajectory is parametrized by a function $f(\Delta\tau)$, according to Eqs.\ (\ref{resprate}) and (\ref{Prop}), the response rate is given by
\be\label{ncresprate}
\dot {\cal F}_\ell=-\frac{i}{ 4\pi^2}\!\int_{i\infty}^{-i\infty}\!\! d \Delta \tau_E \,e^{\Delta E\Delta \tau_E} \left[\frac{1-e^{-f(\Delta \tau_E)/2\ell^2}}{ f(\Delta\tau_E)}\right]\ .
\ee
The corrections due to the minimal length lie only in the damping term in the brackets, and this suggests a suppression of the rate. If $f(\Delta\tau_E)$ is sufficiently smooth, the integrand is holomorphic for all $\Delta\tau_E$, since no singularities occur along the integration path.
If the Jordan lemma is satisfied, the integral vanishes by the Cauchy theorem.

Unfortunately, this does not apply  to a detector moving with acceleration $1/\alpha$ along with a Rindler observer, for which one has, in Euclidean space,
 $f_{\rm R}(\Delta\tau_E)=4\alpha^2 \sin^2\left(\Delta\tau_E/ 2\alpha\right)$. In this case, one can show that the integrand in  (\ref{ncresprate}) is unbounded along straight lines parallel to the real axis \footnote{We thank R.\ Parentani for pointing out this aspect.}.
 However, the integral (\ref{ncresprate}) in the Rindler case is nothing but the Fourier transform of a Gaussian-like function peaked around $\Delta\tau_E =0$. After Wick rotating back the time, we can evaluate the integral by using the saddle point approximation
\bea
\dot {\cal F}_\ell=- \int_{-\infty}^{+\infty}\!\! d \Delta \tau \,e^{-i\Delta E\Delta \tau}\,e^{\,-\ln \{ -1/G_\ell[f_{\rm R}(\Delta\tau)]\} }\ ,
\label{approxrate}
\eea
expanding $\ln (-1/ G_\ell)$ around $\Delta\tau =0$ up to fourth order.
In this way, the integral can be calculated approximatively, and the result is
\bea\label{zeropointrate}
\dot{\cal F}_{\ell}\simeq-\frac{9\,e^{-\Delta E^2\ell^2}}{ 32\pi^{3/2}\ell}+\frac{\ell \,e^{-\Delta E^2\ell^2} }{16\pi^{3/2}\alpha^2}\ ,
\eea
up to subleading terms ${\cal O}(\ell\Delta E^2)$ and ${\cal O}(\ell^3\Delta E^2/\alpha^2)$. We note immediately that
the leading order term does not depend on $\alpha$. So, at first glance, it might appear disturbing that the rate does not vanish
even for an inertial detector, and that it diverges when $\ell \rightarrow 0$. However, a deeper scrutiny reveals that
in the case $\ell\to 0$, this term is simply disregarded since it is equivalent to the (infinite) contribution coming from
the coincidence limit singularity, usually circumvented by the contour of integration. As the leading term of the rate (\ref{zeropointrate}) is negative, it can be interpreted as a dissipation term, because the detector is no longer moving on a smooth differential manifold, but rather on a rough surface endowed with local exponential dampers. In the frame comoving with the detector, this term is simply related to the energy of the field stored in each quantum cell of size $\ell$, which can be though as constituting the fabric of spacetime. Operatively, such a fabric also prevents big quantum leaps $\Delta E$, for
both decays and excitations, due to the exponential form of (\ref{zeropointrate}).

The Unruh effect might still appear at higher orders. In fact, the next-to-leading order term  in (\ref{zeropointrate}) depends
 on the acceleration and it is positive. It also vanishes in the inertial case, $\alpha\to\infty$, and therefore it
represents the actual ``net'' Unruh effect at the temperature $T_\ell= \ell/16\pi^{3/2}\alpha^2$.
  In any case, in order to measure this, one should  ``calibrate'' the accelerated detector, by subtracting  the dissipation term from the observed rate, in a similar way as in Refs.\ \cite{Pepe1,Pepe2}. Thus, we conclude that the net Unruh rate is negligible and the usual thermal distribution disappears completely. It can also be shown that this result holds for the massive case.

This astonishing result is not trivial to interpret from a physical point of view. At first sight, it is not clear how the ``local'' modifications of the UV behavior of the field, according to (\ref{ginp}), can ``globally'' affect the polar singularities, which extend on a infinite  domain of $\Delta\tau$, and probe the IR nature of the field. To clarify this question let us look at the Rindler observer with $\ell = 0$, and consider the following form of the Wightman-Green function for the massless case
\bea
G^{+}(\Delta\tau) =-\frac{1}{4\pi^2}\sum_{k=-\infty}^{\infty}(\Delta\tau+ik\beta)^{-2}  ,
\eea
where $\beta=2\pi\alpha$. This expression clearly shows that the poles in (\ref{resprate}) are ``reflections'' of the singularity at $\Delta\tau=0$, which occur with period $\beta$ along the imaginary axis,  and we know that $\beta$ is nothing but the inverse of the temperature of the system. However, in the presence of the minimal length  $\ell$, there is no singularity at $\Delta\tau=0$ nor at  other periodic points. Therefore
the Planck spectrum disappears.

Another explanation comes from the delocalization caused by the minimal length. The Unruh effect can be explained by showing that the modes associated to the Rindler observer are not analytic at the point where the right and left Rindler wedges meet \cite{BirDav}. Thus, Rindler modes can only be written as a superposition of both positive and negative frequency Minkowski modes, which are analytic on the entire space. When a minimal length is present, one can argue that the ``meeting point'' of the wedges is delocalized and the modes become holomorphic there. So, modes can trespass on the opposite wedges and form a partial superposition over a region of size $\ell$, which is responsible for a tiny, yet non-vanishing flux.

It is interesting to note that certain boundary conditions at the edge of the Rindler wedges, can in fact cancel the Unruh effect even without the presence of a minimal length, as discussed in \cite{Bel1} (see also \cite{Bel2}, where the effect is recovered in the presence of a Bose condensate). These results were also extended to refute the Hawking radiation in \cite{Bel3}.

Our result contrasts sharply the findings of  \cite{Pepe1,Pepe2} and \cite{Paddy}. Here, the momentum space propagator is modified to covariantly introduce a minimal length, and the UV behaviour is very similar to our case. However, the absence of the Gaussian damping term is crucial, especially in \cite{Pepe1,Pepe2}, where it leads to a large departure from the thermal spectrum, unless a calibration procedure is performed. We also add that our calculation avoids the ambiguities found in \cite{Casadio,Nicolaevici}, when expanding the integrand of (\ref{ncresprate}) with respect to $\ell$. Finally, our result is in line with the calculations of \cite{piazza}.

The next logical step would be to clarify whether our result holds only at first order in $\gamma^2$  or if it can be extended to the subleading term $\delta \Psi$. The problem is difficult to address as, in this case, the contributions of the field and of the detector cannot be separated as in Eq.\ (\ref{sepcont}). In fact, the system is no longer governed by a linear Hamiltonian such as  $L_{\mathrm {int}}$. Another way to see this is to realize that a subleading order analysis would require the delocalization of the detector too, whose extension should be at least of the order $\ell$. The same holds for the works \cite{Paddy} and \cite{Pepe3}, where the authors introduce a minimal length but keep the detector  as point-like. Generally speaking, one expects that a detector with a size $L$ will not be able to register quanta of wavelength smaller than $L$, thus providing for a UV cut-off. However, this does not affect the pole structure of the Wightman function, which is at the heart of our results, as well as of the ones of \cite{Paddy} and \cite{Pepe3}. A similar situation occurs in conventional field theory, where the coupling between detector and field does not accord with the uncertainty principle, according to the prescriptions of the semiclassical analysis. In conclusion, we can reasonably expect that if a Planck spectrum can be experimentally observed, its intensity would be at least of order $\gamma^4$.

Concerning the phenomenological implications of our results, we argue that the scenario of the thermalization of Color Glass Condensates (CGC) might be drastically modified \cite{Kharzeev:2005iz}. Recently, it has been proposed that the phase transition from a CGC to a QGP could be driven by the Unruh thermal bath. As partons are subject to huge accelerations in heavy ion collisions, by increasing the energy one can increase the temperature of the thermal bath. Indeed, in strong color fields, partons can have accelerations $1/\alpha\sim 1\ \mathrm{GeV}$, corresponding to a Unruh temperature $T\sim \ 10^{-1}\ \mathrm{GeV}$, an energy that might be sufficient to trigger the transition to the QGP.  According to our findings however, one is left with a low Unruh temperature, which would be $T_\ell\sim 10$ keV for the most optimistic case when $1/\ell\sim 1$ TeV, as in the presence of extra spatial dimensions.  We also expect that the usual thermal bath  could survive at the most at the $\gamma^4$ order. Therefore, the thermalization could be too weak to drive the phase transition.
In other words, a relevant Unruh flux would show up only for accelerations $1/\alpha$ of the order of the fundamental scale $M_*$, whatever it is.
In principle, this argument still holds at the eV scale, but it is physically difficult to believe that atomic physics can be modified by the Planck length. Anyway, for systems like atomic traps, the Unruh bath could yet be used as a ``yes/no tool kit'' to understand the relevance of any intrinsic microscopic scale within the system, other than the Planck length. A suppression of the thermal bath could be interpreted as the signature of  an unknown microscopic scale.

On theoretical grounds, the work presented here can have a strong impact on other effects related to the acceleration radiation. As we mentioned before, the presence of a minimal length is common to many theories, and our findings are valid quite independently of the particular fundamental theory adopted. Thus, we believe that  phenomena such as the Hawking effect and the particle production on time-dependent backgrounds, and their counterparts in analogue models, should all be critically reviewed.


\begin{acknowledgments}
\noindent P.\ N.\  is supported by the Helmholtz International Center for FAIR within the
framework of the LOEWE program (Landesoffensive zur Entwicklung Wissenschaftlich-\"{O}konomischer
Exzellenz) launched by the State of Hesse.  M.\ R.\ is supported by the Swiss National Science Foundation. We wish to thank R.\ Balbinot, R.\ Parentani and E.\ Spallucci for reading the manuscript and providing for very useful suggestions.
\end{acknowledgments}



\begin{thebibliography}{99}

\bibitem{DW}
B. S. DeWitt,
in {\it Gravitation: An Introduction to Current Research},
edited by L. Witten (Wiley, New York, 1962), pp. 266 �- 381.

\bibitem{Akhmedov:2006nd}
  E.~T.~Akhmedov and D.~Singleton,
  Int.\ J.\ Mod.\ Phys.\  A {\bf 22}, 4797 (2007).

\bibitem{Akhmedov:2007xu}
  E.~T.~Akhmedov and D.~Singleton,
  Pisma Zh.\ Eksp.\ Teor.\ Fiz.\  {\bf 86}, 702 (2007).

\bibitem{Kharzeev:2005iz}
  D. Kharzeev and K. Tuchin,
  Nucl. Phys. A{\bf 753}, 316 (2005).

\bibitem{Crispino:2007eb}
 L.~C.~B.~Crispino, A.~Higuchi and G.~E.~A.~Matsas,
  Rev.\ Mod.\ Phys.\  {\bf 80}, 787 (2008).

\bibitem{Barcelo}
  C.~Barcelo, S.~Liberati and M.~Visser,
  Living Rev.\ Rel.\  {\bf 8}, 12 (2005).

\bibitem{Balbinot}
  R.~Balbinot {\it et al.},
  Riv.\ Nuovo Cim.\  {\bf 28}, 1 (2005).

\bibitem{Carusotto:2008ep}
  I.~Carusotto {\it et al.},
  New J.\ Phys.\  {\bf 10}, 103001 (2008).

  \bibitem{rosu}
H.C. Rosu,
Gravitation Cosmol. {\bf 7}, 1 (2001).

\bibitem{Jacob}
  T.~Jacobson and D.~Mattingly,
  Phys.\ Rev.\  D {\bf 64},  024028 (2001).

\bibitem{staro}
  A.~A.~Starobinsky,
  Pisma Zh.\ Eksp.\ Teor.\ Fiz.\  {\bf 73}  415 (2001)
  [JETP Lett.\  {\bf 73}  371 (2001)].

\bibitem{Martin}
  R.~H.~Brandenberger and J.~Martin,
  Int.\ J.\ Mod.\ Phys.\  A {\bf 17}  3663 (2002).

\bibitem{Corley}
  S.~Corley and T.~Jacobson,
  Phys.\ Rev.\  D {\bf 54}  1568 (1996).

\bibitem{Unruh}
  W.~G.~Unruh and R.~Schutzhold,
  Phys.\ Rev.\  D {\bf 71}  024028 (2005).

\bibitem{MaxUnruh}
  M.~Rinaldi,
  Phys.\ Rev.\  D {\bf 77}  124029 (2008).

\bibitem{PaddyDual}
  T.~Padmanabhan,
  Phys.\ Rev.\ Lett.\  {\bf 78}  1854 (1997).
  




\bibitem{Pepe1}
  I.~Agullo  {\it et al.},
    Phys.\ Rev.\  D {\bf 77}, 104034 (2008).


\bibitem{Pepe2}
I.~Agullo  {\it et al.},
  Phys.\ Rev.\  D {\bf 77}  124032 (2008).


\bibitem{Paddy}
 K.~Srinivasan, L.~Sriramkumar and T.~Padmanabhan,
  Phys.\ Rev.\  D {\bf 58} (1998) 044009.

\bibitem{Pepe3}
  I.~Agullo  {\it et al.},
  Phys.\ Rev.\  D {\bf 80} 047503 (2009).



\bibitem{Granot}
A.\ A.\ Abdo  {\it et al.}, Nature {\bf 461} 08574 (2009).



\bibitem{Smailagic:2003yb}
  A. Smailagic and E. Spallucci,
 J. Phys. A{\bf 36}, L467 (2003).


\bibitem{Smailagic:2003rp}
  A. Smailagic and E. Spallucci,
  J. Phys. A{\bf 36}, L517 (2003).

\bibitem{Smailagic:2004yy}
  A.~Smailagic and E.~Spallucci,
  J. Phys. A  {\bf 37}, 1 (2004)
  [Erratum-ibid.  A {\bf 37}, 7169 (2004)].


\bibitem{Nicolini:2005zi}
  P.~Nicolini,
  J. Phys. A  {\bf 38}, L631 (2005).

\bibitem{Nicolini:2005vd}
   P.~Nicolini, A.~Smailagic and E.~Spallucci,
  Phys. Lett.  B {\bf 632}, 547 (2006).

\bibitem{Ansoldi:2006vg}
  S.~Ansoldi {\it et al}.,
  Phys. Lett.  B {\bf 645}, 261 (2007).


\bibitem{Rizzo:2006zb}
  T.~G.~Rizzo,
  JHEP {\bf 0609}, 021 (2006).

\bibitem{Spallucci:2008ez}
  E.~Spallucci, A.~Smailagic and P.~Nicolini,
  Phys. Lett.  B {\bf 670}, 449 (2009).


\bibitem{Nicolini:2009gw}
  P.~Nicolini and E.~Spallucci,
  arXiv:0902.4654 [gr-qc].


\bibitem{Nicolini:2008aj}
  P.~Nicolini,
  Int. J. Mod. Phys.  A {\bf 24}, 1229 (2009).

\bibitem{Rinaldi:2009ba}
  M.~Rinaldi,
  arXiv:0908.1949 [gr-qc].

\bibitem{Spallucci:2006zj}
  E.~Spallucci, A.~Smailagic and P.~Nicolini,
  Phys. Rev.  D {\bf 73}, 084004 (2006).

\bibitem{Casadio:2008qy}
  R.~Casadio and P.~Nicolini,
  JHEP {\bf 0811}, 072 (2008).

\bibitem{BirDav}
  N.~D.~Birrell and P.~C.~W.~Davies, {\it Quantum Fields In Curved Space} (Cambridge University Press, Cambridge, England, 1982)

\bibitem{UnruhEff}
W.~G.~Unruh,
  Phys.\ Rev.\  D {\bf 14}  870 (1976).
  
\bibitem{Bel1}
N.~B.~Narozhnyi, A.~M.~Fedotov, B.~M.~Karnakov, V.~D.~Mur and V.~A.~Belinsky,
  Phys.\ Rev.\  D {\bf 65} (2002) 025004.

\bibitem{Bel2}
V.~A.~Belinskii, N.~B.~Narozhny, A.~M.~Fedotov and V.~D.~Mur,
  Phys.\ Lett.\  A {\bf 331} (2004) 349.

\bibitem{Bel3}

V.~A.~Belinsky,
  Phys.\ Lett.\  A {\bf 354} (2006) 249;
V.~A.~Belinski,
 ``On the Tunnelling Through the Black Hole Horizon,''
  arXiv:0910.3934 [gr-qc].
  
\bibitem{MaxNew}
M.~Rinaldi,
  ``Particle production and transplanckian problem on the non-commutative plane,''
  arXiv:1003.2408 [hep-th].


\bibitem{Casadio}
  R.~Casadio {\it et al}.,
  Phys. Rev.  D {\bf 73}, 044019 (2006).

\bibitem{Nicolaevici}
  N.~Nicolaevici,
  Phys. Rev.  D {\bf 78}, 088501 (2008).

\bibitem{piazza}
  F.~Costa and F.~Piazza,
  arXiv:0805.0806 [hep-th].


\end{thebibliography}
\end{document}